\documentstyle[twocolumn,aps]{revtex}
\input psfig
\begin{document}
\newcommand {\be}{\begin{equation}}
\newcommand {\ee}{\end{equation}}
\newcommand {\bea}{\begin{eqnarray}}
\newcommand {\eea}{\end{eqnarray}}
\newcommand {\nn}{\nonumber}

\draft
\twocolumn[\hsize\textwidth\columnwidth\hsize\csname @twocolumnfalse\endcsname
%
%
%

\title{Spectral Analysis of Correlated One-Dimensional Systems with
Impurities}

\author{Stephan Haas}
\address{Theoretische Physik,
ETH-H\"onggerberg, CH-8093 Zurich, Switzerland}

\date{\today}
\maketitle

\begin{abstract}
An averaging procedure is proposed to account for spectral features of
correlated one-dimensional systems in the presence of non-magnetic
impurities.
The dynamical spin structure factor for a corresponding random ensemble of
Heisenberg chain segments
is calculated by exact numerical diagonalization. It is shown that
a few-pole approximation is sufficient to describe the numerical results.
A similar analysis is proposed for the discussion of 
experimental spectra, such as obtained by inelastic neutron scattering 
measurements on Zn-doped CuO chains.
By examination of the disorder-induced
pseudo-gap, the loss of spectral weight, 
and the discrete peak structures due to smallest-cluster
contributions, the underlying  
impurity distribution function can be determined.
\end{abstract}

\vskip2pc]
\narrowtext

It is known that electrons in one dimension are localized by 
an infinitesimal amount of disorder.\cite{abrahams}
Analogously, their spin degrees of
freedom develop a spectral gap when impurities are placed 
into an infinite chain. 
The branch cut corresponding to the
collective spinon continuum of the pure system splits into a discrete 
set of poles 
characterized by conformal towers of excitations, and 
the spacing between the poles in a given
one-dimensional segment decreases with increasing size.
In this paper it is proposed that
an average over a distribution of one-dimensional segments, 
defined by the regions between randomly placed impurities, should be
taken when calculating physical quantities. At low energies, the finite 
system gaps of all contributing segments then add up to form a pseudo-gap.
By analyzing the peak structure, the overall spectral weight, and the
pseudo-gap features of 
quasi-one-dimensional crystals in the presence of non-magnetic
impurities, such as Zn-doped CuO chains\cite{kim}
and non-stochiometric
$\alpha'$--NaV$_2$O$_5$,\cite{isobe}
the respective distribution function of lengths of
chain segments can thus be deduced, indicating the extent to which
the introduction of
defects has pushed the system into a mesoscopic regime.

As a non-trivial paradigm for correlated one-dimensional systems, let us
thus focus on the antiferromagnetic spin-1/2 Heisenberg chain in the presence 
of non-magnetic impurities. Such a model is realized for example in
Zn-doped SrCuO$_2$,
where the antiferromagnetic superexchange between neighboring Cu$^{2+}$
$\rm d_{x^2-y^2}$ electrons is mediated by the filled O$^{2-}$ p-orbitals.
By substituting Zn$^{2+}$ for Cu$^{2+}$, static vacancies are created,
and the infinite chain is separated into segments of length $l$ which follow
a distribution function $P(l)$. 

The dynamical structure factor S(q,$\omega$)
of the infinite antiferromagnetic spin-1/2 Heisenberg chain
is described approximately
by a two-spinon continuum of excitations which can be probed
by inelastic neutron scattering experiments.\cite{tennant}
It is bounded from below 
by the des Cloiseaux-Pearson dispersion,
$\rm \omega_1(q) = (\pi J/2) |sin(q)|$,\cite{cloiseaux} and from above
by the maximum energy of two unbound spinons $\rm \omega_2(q) = 
\pi J |sin(q/2)|$.\cite{muller} Thus at q=$\pi$ the two spinons have
maximal phase 
space, and the dynamical structure factor
diverges as S($\pi$,$\omega$) $\sim$ 
(1/$\omega$)ln(1/$\omega$) at the
lower boundary.\cite{muller,fledderjohann} For finite systems with an even number of
sites, $l$, the corresponding branch cut splits into a discrete set of
$l$/4 poles for $l$=4,8,12,... , and ($l$+2)/4 poles for $l$=2,6,10,...
(see Fig. 1(a)). Chains with an odd number of sites do not contribute
at q=$\pi$.\cite{footnote1} 
The finite-size scaling behavior of the pole positions, shown in Fig. 1(b),
is well described by the towers of excitations found in conformal field
theory (CFT),
when logarithmic corrections due to marginally relevant 
operators are taken
into account.\cite{affleck,alcaraz,nomura} 
Thus the pole positions are of the form
\bea
\omega_i(l) = \alpha_i/l + \beta_i/(l \ln(l)),
\eea
where $\alpha_i$ and $\beta_i$ are treated as fit parameters. 

\begin{figure}
\centerline{\psfig{figure=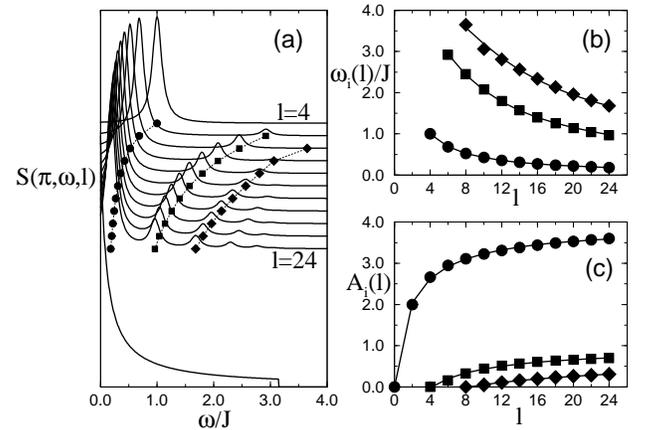,height=6.0cm,angle=0}}
\caption{
(a) Finite-size scaling of the q=$\pi$ dynamical structure
factor of the antiferromagnetic
spin-1/2 Heisenberg chain. In the thermodynamic limit
($l=\infty$) the discrete set of poles merges into a single branch cut.
The $\delta$-peaks have been given a width of $\epsilon$=0.1J.
(b) Finite-size scaling of the first three pole positions in (a).
(c) Finite-size scaling of the corresponding pole weights.
The solid lines in (b) and (c) are fits to the finite cluster data
discussed in the text.
}
\end{figure}

In Fig. 1(b), 
the first three pole positions for finite-size
clusters up to $l$=24 (symbols) are shown along with the fits
to Eq. 1
(solid lines). There is good agreement 
of $\alpha_i$ with the predictions from CFT,\cite{affleck} 
e.g.  $\alpha_1^{CFT} = \pi^2J/2$, while the fit yields
$\alpha_1 = 4.614J \approx 0.94 \alpha_1^{CFT} $.

The pole positions may be calculated directly from the Bethe Ansatz
equations or approximately using CFT, whereas their
respective matrix elements are
accessible only to numerical diagonalizations of small clusters. 
In Fig. 1(c) it is shown that a fit of the finite cluster
pole weights, $A_i(l)$, to the form
\bea
A_i(l) = a_i + b_i \ln(1+l) + c_i \ln(1 + \ln(1+l)),
\eea
gives good results, and
thus permits extrapolation to larger system sizes.
The dynamical spin
structure factor at q=$\pi$ for a system of length $l$ can then
be written in the form
\bea
S(\pi,\omega,l) = \sum_i A_i(l) \delta(\omega - \omega_i(l)),
\eea
where the sum is taken over all poles. In practice, the $\delta$-peaks are
replaced by Lorentzians of width $\epsilon$, which
throughout this paper will be taken as
$\epsilon = 0.1J$.

Let us now turn to the effect of vacancies entering an ideal infinite 
chain i.e. by doping the system with non-magnetic impurities. Even
mobile carriers which can be introduced into the chain by 
out-of-plane substitutions (such as La$^{3+}$ for Sr$^{2+}$ in SrCuO$_2$
\cite{footnote2})
tend to become localized due to
small random potentials, and hence may take the same role as static vacancies.
If the impurities enter the chain in a completely random manner, the
lengths of the chain segments thereby created follow a distribution
function 
\bea 
P(l) = \rho \exp(-\rho l),
\eea
where $\rho$ is the concentration of impurities per lattice site ($0 < 
\rho < 1$).
\cite{laukamp} Note that $P(l)$ is properly normalized, and that its $n^{th}$ 
moment is given by $\langle l^n \rangle = n!/\rho^n$. 
As illustrated in Fig. 2(a), $P(l)$ is
weighted towards smaller segments, with a maximum at 
$P(0) = \rho$. Thus for large impurity concentrations, the average
over chain segments is dominated by contributions from
the smaller clusters. On the other hand, in the dilute limit ($\rho \ll 1$)
$P(l)$ becomes flatter, and thus larger segments contribute appreciably
to the average.

The average dynamical structure factor can consequently be calculated from
\bea
S(\pi,\omega) = \sum_l P(l) S(\pi,\omega,l),
\eea
where the sum is taken over all contributing chain segments with a 
non-vanishing Fourier
component at q=$\pi$. In the case of large impurity concentrations, this
sum may be cut off beyond a certain cluster length $l_{max}$, because the
contributions of segments with $l > l_{max}$ are exponentially suppressed by
$P(l)$. Therefore, as long as $l_{max}$ is smaller than the maximum cluster
which can be studied with present computer capacities ($\leq 36$ for 
Heisenberg systems), very high accuracy can be achieved by simply taking the
sum in Eq. 5 over all available finite systems. However, in the dilute
impurity limit this procedure will necessarily break down, and we propose to
use instead the fits obtained in Eqs. 1 and 2.\cite{footnote3} 

To estimate the quality of such a few-pole approximation, we compare the
spectrum $S(\pi,\omega)$
obtained from cluster diagonalizations (Fig. 2(b)) at a rather large impurity
concentration ($\rho$=0.3), where $l_{max} < 24$ and this procedure is most
reliable, with that of a three-pole approximation using Eqs. 1 and 2 and 
extending the sum in Eq. 5 up to $l = 10000$ (Fig. 2(c)). As can be seen
by comparing the figures, the agreement
between these two methods is very good for large $\rho$: the integrated 
spectral weight differs less than 0.5\% at $\rho$=0.3. However, in the
dilute impurity limit the three-pole approximation does account for the
spectral weight due to the larger segments, yielding for example
14\% more integrated
intensity than the finite
cluster average with $l=4, ..., 24$ at $\rho$=0.1.
In the following,
when discussing the evolution of the dynamical spin
structure factor with impurity
doping, we will thus use the few-pole procedure
rather than the average over system sizes available to exact
diagonalization calculations.

\begin{figure}
\vspace{3mm}
\centerline{\psfig{figure=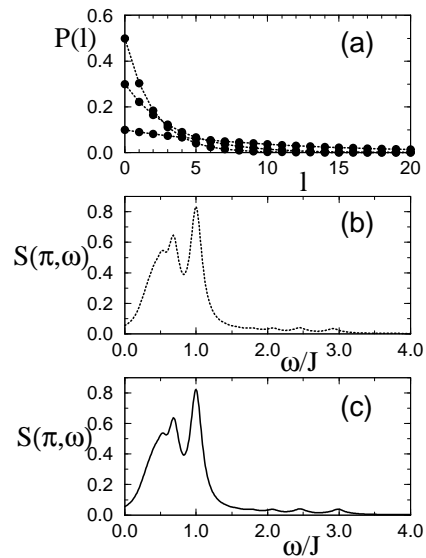,height=7.0cm,angle=0}}
\caption{
(a) Distribution function of chain segments in a one-dimensional
system with randomly placed non-magnetic impurities. $P(l)$ is shown for
impurity concentrations $\rho$ = 0.1, 0.3, and 0.5.
(b) Dynamical structure factor at q=$\pi$ for an antiferromagnetic spin-1/2
Heisenberg chain with $\rho$ = 0.3, calculated from an average over finite
cluster spectra with $l=4, ... , 24$.
(c) Same as (b), but calculated using a three-pole approximation.
}
\end{figure}

In Fig. 3 the evolution of $S(\pi,\omega)$ is shown as a function 
of $\rho$. There are three main features that occur
in $S(\pi,\omega)$ upon randomly introducing vacancies into the chain:
(i) the integrated spectral weight decreases rapidly with 
increasing $\rho$, (ii) a pseudo-gap develops at small frequencies,
(iii) at larger frequencies (of order $J$) a discrete peak structure
emerges,
dominated by the contributions of the smallest cluster segments. 

In order to discuss the low-energy features (i) and (ii), it is useful 
to consider the continuum limit of Eq. 5 since they are dominated by the
contributions of the
larger chain segments. We emphasize, however, that the discreteness of the
peak structure at higher energies
- which we consider to be a major characteristic -
cannot be accounted for in this limit. For simplicity
let us consider the single-pole approximation to leading order, i.e. 
$\omega_1(l) = \alpha_1/l$ and $A_1(l) = b_1 \ln(1+l)$ (for the 
weight of the lowest pole: $a_1 =0$). Replacing the sum in Eq. 5 with an
integral it is then found that
\bea
S(\pi,\omega) \approx
\frac{b_1 \alpha_1 \rho}{\omega^2} \ln(1 + \frac{\alpha_1}{\omega})
\exp(\frac{- \alpha_1 \rho}{\omega}), \ \ \ \  (\omega\ll J).
\eea

Hence at low energies $S(\pi,\omega)$ is exponentially suppressed, and the
peak of the corresponding pseudo-gap feature 
moves towards larger frequencies with
increasing $\rho$. The frequency sum-rule in the single-pole approximation
yields $S(\pi) = \int {\rm d}\omega S(\pi,\omega) \approx b_1 \exp(\rho)
{\rm E}_1(\rho)$, where E$_1(\rho)$ is the first exponential integral. This
result is in good agreement with the $\rho$-dependence
of the frequency-integrated spectral weight 
obtained within the full three-pole approximation shown in Fig. 4(a).

\begin{figure}
\vspace{3mm}
\centerline{\psfig{figure=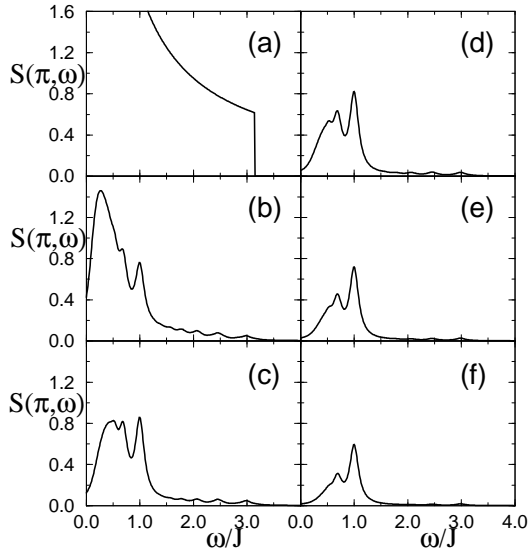,height=7.0cm,angle=0}}
\caption{
Evolution of $S(\pi,\omega)$ in the antiferromagnetic spin-1/2
Heisenberg chain as a function of impurity concentration, calculated
from a three-pole approximation. In
(a)-(f) the respective spectra are shown for $\rho$=0.0, 0.1, 0.2,
0.3, 0.4, and 0.5.
}
\end{figure}

From Eq. 6 it is seen that, strictly speaking, there should be no spectral 
weight at $\omega = 0$. However, if the $\delta$-peaks in Eq. 3
are given a finite lifetime, 
i.e. by replacing them with Lorentzians of width $\epsilon$ in order to mimic
the experimental situation, there will be some residual
spinon density of states $S(\pi,0)$. As shown in Fig. 4(b), this quantity 
is exponentially suppressed with increasing $\rho$, and it diverges in the 
dilute impurity limit.

According to our interpretation of the pseudo-gap by taking the 
continuum limit (Eq. 6), it arises due to the exponentially suppressed
contributions of the largest segments (corresponding to the smallest
finite-size gaps) to the average in Eq. 5. On the other hand, at 
higher energies ($\omega \approx J$) the discrete nature of the smaller
contributing clusters should be visible, with the largest features 
coming from the smallest clusters. Clearly, the lowest pole of the
smallest contributing cluster at $\omega = J$
is well defined in Figs. 3(b)-(f),
advancing into the dominating feature of the spectrum for large $\rho$.
At smaller impurity concentrations (Figs. 3 (b)-(d)) the lowest pole of 
the second smallest cluster is also well separated from the low-energy
continuum.
At larger $\rho$, however, it merges into the low-energy
pseudo-branch. Summarizing the results for $S(\pi,\omega)$ in the presence
of randomly placed static vacancies, an inelastic neutron 
scattering experiment on such
a system should observe a marked decrease of spectral weight with increasing
impurity concentration, a low-energy pseudo-gap feature, and - resolution
permitting - some split-off ``high-energy" peaks due to (and indicating the
size of) the smallest segments contributing to the average.
 
\begin{figure}
\vspace{-18mm}
\centerline{\psfig{figure=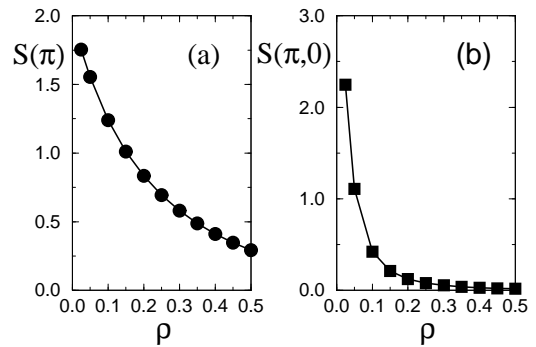,height=7.0cm,angle=0}}
\vspace{-5mm}
\caption{
(a) Dependence of the frequency-integrated dynamical structure factor
at wavevector $\pi$ on the concentration of non-magnetic impurities.
(b) Residual spinon density of states as a function of impurity 
concentration.
}
\end{figure}

In real materials, impurities do not necessarily enter a chain in an
entirely uncorrelated manner, but may e.g. due to a 
lattice potential commensurate
with the surrounding crystal structure favor certain distances 
between impurity sites. Effects of this kind can be accounted for
by refining the probability distribution function in the above analysis.
E.g. in the presence of a single preferred segment size $l_0$ one
finds 
\bea
P(l) = \frac{\rho}{2 - \exp(-\rho l_o)} \exp(-|l - l_0|\rho),
\eea
which is peaked at $l = l_0$, and hence weighs the contribution by 
segments of length $l_0$ most highly in the average. A spectral analysis
of inelastic neutron 
scattering experiments on impurity doped systems can thus indicate the
underlying distribution function of chain segments. In particular, at large
$\rho$ the characteristic ``high-energy" features coming from the
segments with length $l_0$ should dominate the spectral weight.

Odd-site segments do not contribute to the dynamical structure factor
at $q = \pi$, but they have spectral weight 
at other momentum transfers which are contained in their
Fourier spectrum. 
While even-site segments have a singlet
ground state associated with a gap in their excitation spectrum,
odd-site segments have a doublet ground state, and the effective low-energy 
spin-1/2 degree of freedom in these clusters leads to a single-spinon
contribution to $S(q,\omega)$ at small frequencies.         
This low-energy spectral weight from wavevectors close to $\pi$ may    
``spill over", causing a slight enhancement of the zero-frequency spinon density
of states $S(\pi,0)$.\cite{footnote4} 

If there are residual inter-chain exchange couplings $J_\perp$
( $ |J_\perp| \ll |J|$) 
in a quasi-one-dimensional
magnetic material, there is a transition to a three-dimensional
ordered phase at a low temperature $T_N \sim |J_\perp|$.\cite{schulz}
However, in the presence of an impurity-induced
pseudo-gap in the chain spectrum,
$S(\pi,0)$ does not diverge anymore at low temperatures, and thus 
$T_N $ is decreased with increasing $\rho$.
In the case of a vanishing residual spinon density of states
this transition is completely suppressed, as it is 
the case for example in weakly coupled two-leg
Heisenberg ladders.\cite{azuma}
In this sense impurities may have a stabilizing
effect on one-dimensional phases.\cite{footnote5}

As a consequence of the spin-charge separation
in one-dimensional systems away from half-filling, 
the spin and charge degrees of freedom can be treated equivalently.        
Therefore it can be expected for a segmented chain with 
(non-localized) mobile charge carriers
that the results obtained above for $S(q,\omega)$
hold also for the dynamical charge response $N(q,\omega)$.

To conclude, we have proposed an averaging procedure to account for the
spectral features of one-dimensional systems in the presence of 
non-magnetic impurities. At small doping concentrations, 
the contributions of the longer segments become increasingly important,
and the
average over their lowest poles yields a pseudo-gap feature
in the spin excitation spectrum at low 
frequencies. At larger
impurity concentrations, the spectrum is dominated by discrete peaks at $\omega
\sim J$ contributed by the shortest segments.

When static vacancies are present in a correlated one-dimensional
system, the formation of a pseudo-gap as discussed above will compete 
or coexist with
other low-energy features, 
such as gaps due to frustrating exchange interactions
or
spin-Peierls instabilities.\cite{uhrig,haas,ain,arai}
In such systems it then remains to decide how much of the 
observed discrete
low-frequency peak structure is due to such bulk instabilities
or maybe to finite segment contributions.\cite{footnote6}

We wish to thank 
A. V. Balatsky,
E. Dagotto, N. Furukawa, M. Laukamp, F. Mila, T. M. Rice, and  M. Sigrist
for useful discussions,
and acknowledge
the Swiss National Science Foundation for financial support.

%
%


\begin{references}
\bibitem{footnote1}Note that the spectra shown in Fig. 1(a) were obtained on
chains with periodic boundary conditions and an even number of
sites. However, a Fourier analysis of the equivalent systems with
open boundary conditions yields similar spectra with the same scaling 
behavior.
\bibitem{footnote2}La$^{3+}$ substitution for Sr$^{2+}$
would actually donate an additional
electron to the system. However, the electron could combine with one of the
Cu$^{2+}$ to form a singlet which acts as an effective hole in the chain.
\bibitem{footnote3}Although only a finite number of poles may be included
in such a calculation, the dominant, low-energy contributions arise from
the lowest few poles, and are thus accounted for to high accuracy.
\bibitem{footnote4}Odd-site segments also contribute multi-spinon spectral
weight at wavevectors close to $\pi$ and frequencies $\omega \sim O(1/l)$. 
These contributions can be treated analogously to those of the even-site
segments.
\bibitem{footnote5}At all impurity concentrations,
the low-energy contributions
to the spectral weight of the odd-site segments may lead 
to a small but finite $T_N$.
\bibitem{footnote6}The discrepancy of the broad measured static spin
structure factor $S(q)$ in $\rm CuGeO_3$\cite{arai} and theoretical predictions
\cite{muller,haas} may originate in a random segmentation of the
idealy infinite $\rm CuO_2$ chains in this material.
\end{references}
\end{document}